 \documentclass[preprint2]{emulateapj-rtx4}

\usepackage{txfonts} \usepackage{epsfig} \usepackage{graphicx}
\usepackage{natbib} 
\begin{document}

\title{A 10,000 Years Old Explosion in DR21}

\shortauthors{Zapata, et al.}

\author{Luis A. Zapata\altaffilmark{1}, Johannes
  Schmid-Burgk\altaffilmark{2}, Nadia P\'erez-Goytia\altaffilmark{1},
  Paul T. P. Ho\altaffilmark{3,4}, \\ Luis F. Rodr\'\i guez\altaffilmark{1},
  Laurent Loinard\altaffilmark{1}, and Irene Cruz-Gonz\'alez\altaffilmark{5} }

\altaffiltext{1}{Centro de Radioastronom\'\i a y Astrof\'\i sica,
  UNAM, Apdo. Postal 3-72 (Xangari), 58089 Morelia, Michoac\'an,
  M\'exico} \altaffiltext{2}{Max-Planck-Institut f\"{u}r
  Radioastronomie, Auf dem H\"ugel 69, 53121, Bonn, Germany}
\altaffiltext{3}{Academia Sinica Institute of Astronomy and
  Astrophysics, Taipei, Taiwan} \altaffiltext{4}{Harvard-Smithsonian
  Center for Astrophysics, 60 Garden Street, Cambridge, MA 02138, USA}
\altaffiltext{5}{Instituto de Astronom\'\i a, Universidad Nacional
  Aut\'onoma de M\'exico, Ap. 70-264, 04510 DF, M\'exico}

\begin{abstract} 
Sensitive high angular resolution ($\sim$ 2$''$) CO(2-1) line
observations made with the Submillimeter Array (SMA) of the flow emanating
from the high-mass star forming region DR21 located in the Cygnus X
molecular cloud are presented. These new interferometric observations
indicate that this well known enigmatic outflow appears to have been
produced by an explosive event that took place about 10,000 years
ago, and that might be related with the disintegration of a massive stellar
system, as the one that occurred in Orion BN/KL 500 years ago, 
but about 20 times more energetic. 
This result therefore argues in favor of the idea that the
disintegration of young stellar systems perhaps is a frequent phenomenon
present during the formation of the massive stars.
However, many more theoretical and observational studies 
are still needed to confirm our hypothesis.
 
\end{abstract}


\section{Introduction}

Located in the Cygnus X complex at a distance of 1.36 $\pm$ 0.12 kpc
\citep{Ryg2012}, the outflow in DR21 is probably one of the most
enigmatic high-velocity molecular outflows associated with star
forming regions. This outflow is extremely massive ($>$ 3000
M$_\odot$) and energetic ($>$ 10$^{48}$ erg) with a luminosity in the
2 $\mu$m band alone calculated to be 1800 L$_\odot$
\citep{gar1991,gar1992}. The DR21 outflow has been suggested to be
energized by a massive protostar with bolometric luminosity of about
10$^{5-6}$ L$_\odot$ \citep{gar1991,gar1992}, which would exceed the
total bolometric luminosity of DR21 region \citep[10$^{5}$
  L$_\odot$;][]{har1977}.  However, such a massive star has so far
never been clearly found out \citep{cru2007}.

The outflow appears to be bipolar with its strongest molecular lobes
extending northeast-southwest, and emanating from a dense dusty
filament that extends from north to south \citep{sch2010}. However,
there are some faint H$_2$ filaments that show different orientations 
with some of them pointing back to the DR21 main. 

It has been suggested that the axis of this bipolar outflow
might be located very close to the plane of the sky \citep[e.g.][]{gar1991}, 
however, the high velocity wings observed in H$_2$ 
and CO ($-$80.82 to $+$46.84 km s$^{-1}$) and
the overlapping of blue/redshifted emission on both lobes
\citep{cru2007,sch2010} indicate a different scenario for its
orientation.  Commonly bipolar outflows with their axes located 
close to the plane of the sky show slow radial velocities 
with their blue/redshifted lobes well separated  \citep[e.g.][]{pech2012}. 
However, overlapping blue/redshifted emission on 
both lobes might result from an outflow in the plane of the sky 
(that seems to be unlikely due to its high velocity wings) if the outflow 
is thought of as a cone with one side of the cone coming toward 
us and the other side moving away.

\cite{gar1991, gar1992, cyg2003, dav2007} additionally
reported the presence of some collimated flows emanating from DR21
with different orientations to the main NE-SW DR21 outflow. 

In high-density protostar clusters, envelopes and disks provide a 
viscous medium that can dissipate the kinetic energy of passing stars, 
greatly enhancing the probability of capture. This process opens the
possibility of having protostellar mergers that may generate impulsive 
wide-angle outflows, shock-induced masers, radio continuum emission,
and runaway massive protostars. Both the ejection of massive stars 
and the launch of the impulsive outflow may be the result of the 
dynamic interaction and rearrangement of a system of massive stars.
This process has been extensively discussed in \cite{ball2005,ball2011}.   

Very recent sensitive $^{12}$CO(2-1) millimeter and infrared observations 
have suggested that the massive (10 M$_\odot$) and energetic ($\sim$
10$^{47}$ erg) outflow located in the Orion BN/KL region appears to have been
produced by such a violent explosion during the disruption of the
massive young stellar system of which the infrared and radio sources
{\it BN}, {\it n}, and {\it I} were all members about 500 years ago
\citep{zap2009, zap2011a, zap2011b, ball2011}.  This suggests that
the complex of CO and H$_2$ emission in BN/KL is due to a phenomenon
different from the well known collimated outflows produced by the young stars
commonly associated with star formation.  With this recent result in
mind, one could consider if the flow emerging from DR21 could be a
similar case as that occurred in BN/KL 500 years ago. 

We thus here report new SMA interferometric observations of this outflow
trying to elucidate its nature.

\begin{figure*}
\centering \includegraphics[scale=0.46, angle=0]{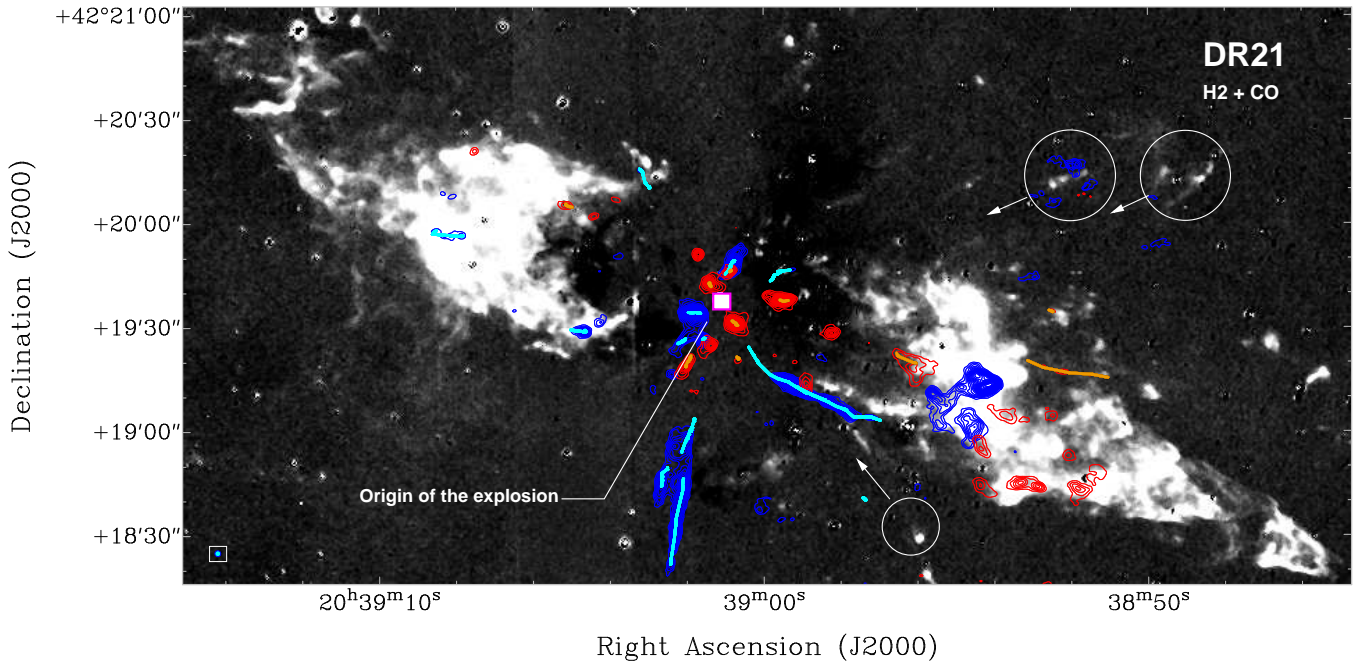}
\caption{ {\scriptsize High velocity redshifted (red/brown) and blueshifted (blue) CO(2-1)
    filaments in the DR21 outflow as observed with the SMA, overlaid
    on an H$_2$ image from Davis et al. (2007). The blue contours are from 13$\%$
    to 90$\%$ in steps of 3$\%$, the peak of the molecular emission; the peak 
    emission is 21 Jy beam$^{-1}$ km s$^{-1}$. The red contours are from 12$\%$
    to 90$\%$ in steps of 5$\%$, the peak of the molecular emission; the peak 
    emission is 16 Jy beam$^{-1}$ km s$^{-1}$.
    Each line represents
    one sequence of positions at which the CO emission peaks in consistent
    velocity channels. Most of the filaments point approximately toward the same
    central position (represented here by a box)  
    $\alpha$=20$^h$ 39$^m$ 1.1$^s$ $\pm$ 5$''$ and
    $\delta$=$+$42$^\circ$ 19$'$ 37.9$''$ $\pm$ 5$''$. 
    The size of the box represents the 
    errors given in the text. There are also some H$_2$ faint
    filaments that appear to point toward the same central position
    \citep{dav2007}. These H$_2$ filaments are marked with white
    circles.  We used the mosaicing mode to cover the entire DR21
    outflow as far as it has been mapped in H$_2$.
    The synthesized beam is shown in the lower left corner. 
    }}
\label{fig1}
\end{figure*}

\section{Observations}

The observations were made with the Submillimeter Array\footnote{The
  Submillimeter Array is a joint project between the Smithsonian
  Astrophysical Observatory and the Academia Sinica Institute of
  Astronomy and Astrophysics, and is funded by the Smithsonian
  Institution and the Academia Sinica.} (SMA) during 2011 August and
2012 July/August. The SMA was in its subcompact, compact, and extended
configurations with baselines ranging in projected length from 7 to
160 m.  We used the mosaicing mode with half-power point spacing
between field centers and covered the entire DR21 outflow mapped in
H$_2$ by \cite{dav2007}, see Figure 1. We concatenated the three 
data sets using the task in MIRIAD called "{\it uvcat}''. The three 
different observations were identical, and only the antenna configuration
of the SMA changed. 

The SMA correlator was configured in 24 spectral ``chunks'' (or
windows) of 104 MHz each, with 128 channels distributed over each
spectral window, providing a spectral resolution of 0.8125 MHz (1.05
km s$^{-1}$) per channel.  The receivers were tuned to a frequency of
230.5387 GHz in the upper sideband (USB), while the lower sideband
(LSB) was centered on 220.5387 GHz. The $^{12}$CO(2-1) transition
was detected in the USB at frequencies close to a frequency of 230.5
GHz. The full bandwidth of the SMA correlator is 8 GHz (4 GHz in each
band).

The zenith opacity ($\tau_{230 GHz}$) was $\sim$ 0.1 -- 0.3,
indicating reasonable weather conditions.  Observations of Uranus
provided the absolute scale for the flux density calibration.  Phase
and amplitude calibrators were the star MWC349a, and the quasar J2007$+$404.
Further technical descriptions of the SMA and its calibration schemes
can be found in \citet{ho2004}.

The data were calibrated using the IDL superset MIR, originally
developed for the Owens Valley Radio Observatory
\citep{Scovilleetal1993} and adapted for the SMA.\footnote{The MIR
  cookbook by C.  Qi can be found at
  http://cfa-www.harvard.edu/$\sim$cqi/mircook.html} The calibrated
data were imaged and analyzed in a standard manner using the MIRIAD, and
KARMA \citep{goo1996} packages. The line image rms noise was around
100 mJy beam$^{-1}$ for each velocity channel at an angular resolution
of $2\rlap.{''}17$ $\times$ $1\rlap.{''}89$ with a P.A. =
$+$74.0$^\circ$.

\section{Results and discussion}

In Figure 1, we show the $^{12}$CO(2-1) molecular emission detected
in our SMA mosaic towards the DR21 region. We construct this large scale integrated 
intensity map only using the emission arising from the high velocity gas and thus avoiding 
the emission in the range from $-$15 to 15 km s$^{-1}$.    
In this velocity range the line emission is spatially extended,
and could not be well sampled by the SMA.   
In our channel velocity maps, we found that the redshifted $^{12}$CO(2-1) 
emission shows radial velocities from $+$3 up to $+$50 km s$^{-1}$, 
while the blueshifted one from $+$3 down to $-$80 km s$^{-1}$. 
This velocity range is in very good agreement to that previously observed by \citet{gar1991, sch2010}. However, our high angular resolution observations revealed clear molecular filaments with different orientations, and some of them well correlated with H$_2$ emission (Figure 1).   

\begin{figure*}
\begin{center}
\includegraphics[scale=0.29, angle=0]{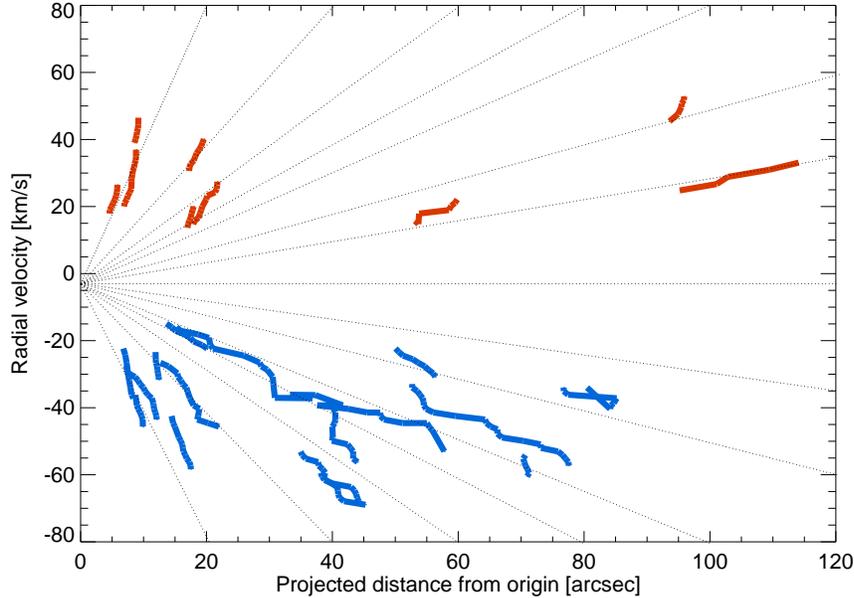}
\caption{Position-velocity diagram of the jet-like CO complex: Radial
  velocity as function of on-the-sky distance from the common center
  for each of the 25 CO(2-1) filaments, with blueshifted structures shown
  in blue, and the redshifted ones in red.  Nearly all filaments seem
  to start from the same common radial velocity of about $-$3 km
  s$^{-1}$ (the small black circle). Black gradient dashed-lines start
  from $-$3.0 km s$^{-1}$ . Note that velocities between $-$15 and $+$15
  km s$^{-1}$ could not be studied because the interferometer cannot
  sample well the extended molecular gas. }
\end{center}
\label{fig2}
\end{figure*}

The CO filaments reported here are composed of about three hundred compact 
features observed in our channel velocity cube with velocity 
windows of 1 km s$^{-1}$ and their positions on the sky are shown in our Figure 1. 
The observational parameters of these 
compact features were found using the task {\it imfit} of MIRIAD.
Each of these filaments show clear velocity gradients that follow a Hubble law (Figure 2).  
Furthermore, most of these filaments appear to point out a common velocity center located 
at about $-$3.0 km s$^{-1}$, in very good agreement with the ambient molecular velocity 
of DR21 \citep[$-$3.0 km s$^{-1}$;][]{sch2010}.

Many of these physical features are also found in the molecular outflow located in BN/KL
and are related to an explosive event occurred 500 years ago as described 
in the introduction. However, the outflow in DR21 has a clear east-west bipolar
shape observed on the H$_2$ maps.  This H$_2$ bipolar shape could be
due to the highly extinction produced by the north-south dusty
lane where many of the CO filaments are revealed for the first time. 
This also seems to be the case in the outflow located in BN/KL, 
see \citet{zap2011a}.  

Using a similar approach to that  applied in \citet{zap2009},  
we determine the position of the origin of the CO filaments
in $\alpha$(J2000)=20$^h$ 39$^m$ 1.1$^s$ $\pm$ 0.3$^s$ and
$\delta$(2000)=$+$42$^\circ$ 19$'$ 37.9$''$ $\pm$ 5$''$. 

This position coincides well with the center of an expanding cometary
HII region (see Figure 3) imaged by \cite{cyg2003}. This positional
coincidence suggests that both the explosive outflow and the large
cometary HII region were probably produced by the same
mechanism. Moreover, at the position of the outflow origin there is
not a single young (radio, submillimeter, or infrared) star,
suggesting that maybe its "source'' is no longer located there, as in
the case of the explosive outflow in Orion BN/KL \citep{zap2009}. 
There is also a depression of centimeter continuum emission at this position 
(see Figure 3).

The explosive Becklin-Neugebauer (BN)/Kleinman-Low (KL) outflow emerging
from OMC1 behind the Orion Nebula was powered by the dynamical decay
of a non-hierarchical multiple system $\sim$ 500 years ago that
ejected the massive stars I, BN, and source n, with velocities of
about 10 -- 30 km s$^{-1}$ \citep{zap2009, zap2011a, zap2011b,
  ball2011}.  In this dynamical decay of a non-hierarchical multiple
system a kinetic energy of about 10$^{47}$ ergs was liberated.

If we assume that the outflow in DR21 was produced by a dynamical decay 
of a non-hierarchical multiple system, the kinetic energy
liberated on that disruption was about 20 times larger
than that in the Orion BN/KL system \citep[$>$ 2 $\times$ 10$^{48}$ erg;][]{gar1991,gar1992}.

Again, very 
crudely assuming that the velocities of
the stars ejected in DR21 have similar velocities as the ones located
in Orion BN/KL (this without any evidence), and a dynamical age of
about 10,000 years\footnote{This is an approximate 1st order estimation as in
  the case of the Orion BN/KL \citep{zap2009}} (obtained from our molecular data, t$_d$
$\sim$ 1.2 $\times$ 10$^{18}$ cm / 4 $\times$ 10$^{6}$ cm s$^{-1}$
$\sim$ 10, 000 yrs) for the outflow in DR21. The reader should note that 
this estimation is also an approximation, since the values for the velocity  
are only radial velocities, and the displacements are only the projected 
components.

We thus found that the ejected
stars should be located about 25$''$ far from the center of the DR21
outflow. This is exactly the distance of the bright infrared source
``DR21-D'' associated with a small cometary HII region (Figure 3) with respect
to the origin of the outflow in DR21.  This suggests that maybe DR21-D
is related with the explosive event in DR21.  
The Figure 3 also reveals a reflecting east-west cavity associated with
the large cometary HII region.  These infrared nebulosities are also
observed in Orion BN/KL and most of them are not heated by stars.
They are possibly heated by strong shocks produced by the explosive
outflow \citep{men1995,zap2011a}.

\begin{figure*}
\begin{center}
\includegraphics[scale=0.35, angle=0]{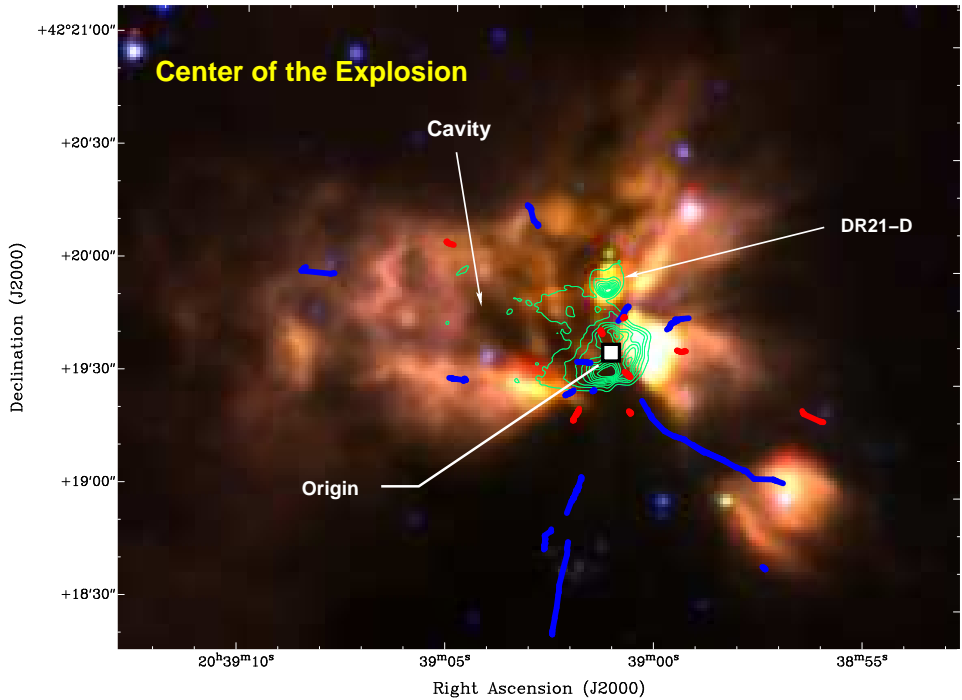}
\caption{ \scriptsize Redshifted and blueshifted CO filaments in the
  DR21 outflow as observed with the SMA, overlaid on a Spitzer
  infrared color image and a VLA 6 cm. emission contour image. In the
  Spitzer infrared color image composite the red represents the 8.0
  $\mu$m, green the 5.8 $\mu$m, and blue the 3.6 $\mu$m emission. The contours
  are 5\% to 90\% with steps of 10\% the peak of the centimeter
  emission; The peak of the centimeter image is 0.13 Jy Beam$^{-1}$.
  The center of the explosive molecular outflow is marked with a white
  square. Note that this center coincides with the center of the compact
  HII region revealed by the 6.0 cm. continuum emission map.
  Additionally, the infrared map reveals a continuum source associated
  with the cometary HII region called DR21-D. Note that the white square
  does not lie at the peak of the 6.0 cm. emission, rather it is located
  in the middle of the HII region, where no emission is detected at all. }
\end{center}
\label{fig3}
\end{figure*}

All the characteristics found in these observations in the outflow located in DR21 
are very similar to the ones reported in the explosive outflow located
in Orion BN/KL \citep{zap2009}. 
We thus propose that the complex of CO and H$_2$ emission located in
DR21 might have been produced by an explosive phenomenon different from the
standard disk-outflows associated with star formation. This outflow
was probably originated during the disintegration of a massive stellar
system as the one occurred in Orion BN-KL some 500 years ago. As mentioned in the
Introduction section,  the process of that disintegration of the stellar system has 
been discussed in detail by \cite{ball2005}.  
The presence of explosive phenomena in two of the
best studied regions of massive star formation suggests that the
phenomenon may be more common than previously thought and that future
detailed studies may reveal more of them in other regions.

\section{Conclusions}

Sensitive high angular resolution ($\sim$ 2$''$) CO(2-1) line
observations made with the Submillimeter Array of the flow emanating
from the massive star forming region DR21 located in the Cygnus X
molecular cloud have been presented. We found about 25 molecular 
blueshifted and redshifted filaments. 
These molecular filaments follow nearly straight lines
and appear to all point toward a common center.  
The radial velocity along each filament follows clearly a Hubble law,
indicative  of an explosive event.
Our observations suggest that this outflow appears to have
been produced by a violent explosion that took place about 10,000
years ago, and that seems to be related with the disintegration of a
massive non-hierarchical stellar system, as the one that occurred in Orion BN/KL, 
but about 20 times more energetic. However, many more theoretical and observational studies 
are still needed to confirm our hypothesis.

\acknowledgments L.A.Z, L. L. and L. F. R. acknowledge the financial
support from DGAPA, UNAM, and CONACyT, M\'exico.  L. L. is indebted to
the Alexander von Humboldt Stiftung for financial support. We are very
grateful to Chris Davis for having provided us with his H$_2$ image of
the DR21 outflow.  This study makes use of Spitzer telescope infrared
images from the archive.  We are very thankful for the suggestions of
anonymous referee that helped to improve our manuscript.

\end{document}